\documentclass[%
 aip,
 amsmath,amssymb,
 reprint,%
]{revtex4-1}

\usepackage{graphicx}% Include figure files
\usepackage{epstopdf}
\usepackage{color}
\usepackage{bm}
\usepackage{amsmath}
\usepackage{changes}

\definecolor{pink}{rgb}{1,1,0} % color values Red, Green, Blue
\definecolor{red}{rgb}{1,0,0}
\definecolor{yellow}{rgb}{1,1,0}
\definecolor{orange}{rgb}{1,0.5,0}
\definecolor{green}{rgb}{0,1,0}
\definecolor{blue}{rgb}{0,0,1}
\definecolor{white}{rgb}{1,1,1}
\definecolor{purple}{rgb}{0.5,0,0.5}

\usepackage{dcolumn}% Align table columns on decimal point
\usepackage{bm}% bold math
\usepackage[mathlines]{lineno}% Enable numbering of text and display math
\usepackage[utf8]{inputenc}
\usepackage[T1]{fontenc}
\usepackage{mathptmx}
\usepackage{etoolbox}
 \usepackage{braket}

\makeatletter
\def\@email#1#2{%
 \endgroup
 \patchcmd{\titleblock@produce}
  {\frontmatter@RRAPformat}
  {\frontmatter@RRAPformat{\produce@RRAP{*#1\href{mailto:#2}{#2}}}\frontmatter@RRAPformat}
  {}{}
}%
\makeatother
\begin{document}

\preprint{AIP/123-QED}

\title{Spin model for the Honeycomb $\rm NiPS_3$}
% Force line breaks with \\
\author{Paula Mellado}%
 \email{paula.mellado@uai.cl}
\affiliation{ 
School of Engineering and Sciences, 
	Universidad Adolfo Ib{\'a}{\~n}ez,
	Santiago, Chile%\\This line break forced with \textbackslash\textbackslash
}%

\date{\today}% It is always \today, today,

\begin{abstract}
In the Van der Waal material $\rm NiPS_3$, Ni atoms have spin S=1 and realize a honeycomb lattice. Six sulfur atoms surround each Ni and split their d manifold into three filled and two unfilled bands. Aimed to determine the spin Hamiltonian of $\rm NiPS_3$, we study its exchange mechanisms using a two-band half-filled Hubbard model. Hopping between d orbitals is mediated by p orbitals of sulfur and gives rise to bilinear and biquadratic spin couplings in the limit of strong electronic correlations. The microscopic model exposed a ferromagnetic  biquadratic spin interaction $\rm K_1$ allowing the completion of a minimal $\rm J_1-J_3-K_1$ spin Hamiltonian for $\rm NiPS_3$. In bulk, a ferromagnetic first nearest neighbor $\rm J_1$ and a more significant antiferromagnetic third nearest neighbor spin coupling $\rm J_3$ agreed with the literature, while in monolayer $\rm J_1$ is positive and very small in comparison. Using a variational scheme we found that a zig-zag antiferromagnetic order is the ground state of bulk samples. The zig-zag pattern is adjacent to commensurate and incommensurate spin spirals, which could hint at the puzzling results reported in  $\rm NiPS_3$ monolayers. 
\end{abstract}
\maketitle
\begin{figure}
\includegraphics[width=\columnwidth]{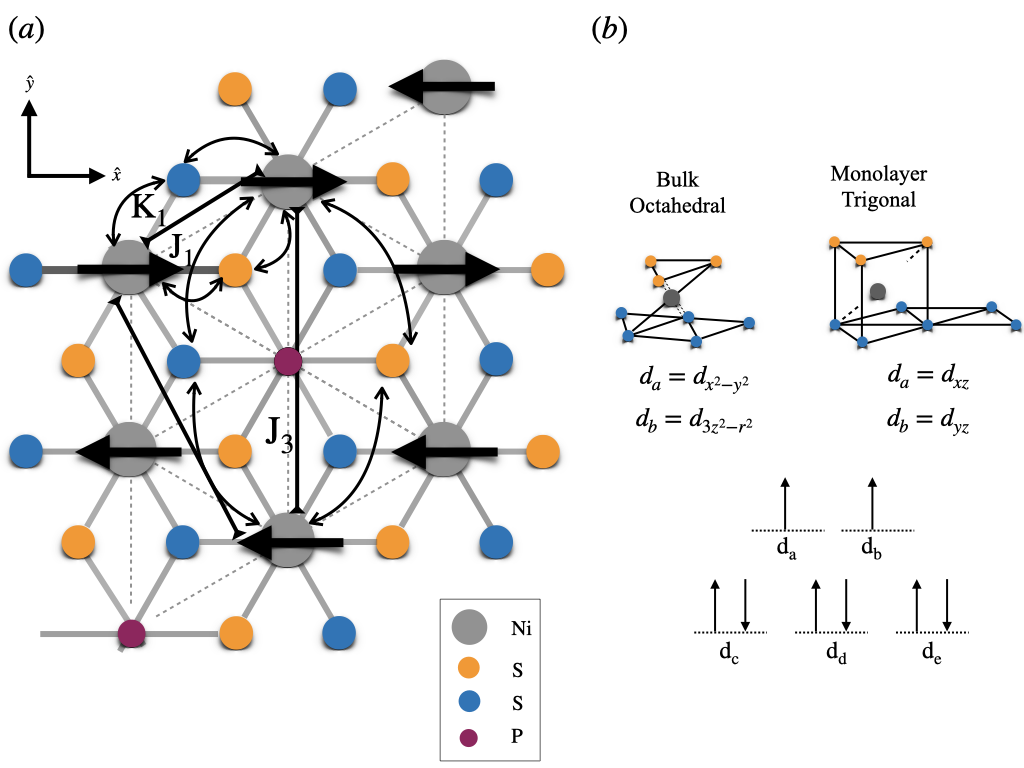}
    \caption{(a) Top view of a single layer of $\rm NiPS_3$. Grey, blue, orange, and cyan-filled circles illustrate Ni, top Sulfur, bottom Sulfur, and P atoms. Black arrows on top of grey circles illustrate the direction of spins in Ni atoms when the system is in the zig-zag magnetic state.   Curved arrows highlight superexchange between Ni atoms mediated by sulfur ions. $\rm J_1$, $\rm J_3$, and $\rm K_1$ denote the first and third nearest neighbor and biquadratic first nearest neighbor spin couplings, respectively. (b) Crystal field in bulk and monolayer and the splitting of the d band \cite{sugano2012multiplets,koch2012exchange}.}
    \label{fig:f1}
\end{figure}

Van der Waal compounds, particularly transition-metal thiophosphates \cite{burch2018magnetism}, are an exciting class of materials. Their negligible interlayer coupling reduces their dimensionality and promotes intriguing electronic and optical quantum effects \cite{seifert2022ultrafast,kim2018charge,gu2019ni,rosenblum1994two,rosenblum1999resonant,basnet2021highly, afanasiev2021controlling,belvin2021exciton,kang2020coherent,ergeccen2022magnetically,jiang2021recent,mattis2012theory,kim2023microscopic,wilson1969transition} while allowing for easy tunability of magnetic exchange and anisotropy through ligand substitution \cite{basnet2022controlling, mak2019probing}. Family members of transition-metal thiophosphates Van der Waal materials have monoclinic space group $\rm C/2m$, with the transition metal atoms forming a planar honeycomb lattice in the ab planes \cite{lanccon2018magnetic}, Fig.\ref{fig:f1}(a). The metal atoms are enclosed in octahedra formed by sulfur atoms and have a phosphorus doublet at the center of the honeycomb hexagons \cite{kertesz1984octahedral,hempel1981ground}. 
 Magnetic susceptibility measurements on single crystals reveal that the family member $\rm NiPS_3$ has the smallest spin S = 1 \cite{kim2021magnetic} and the largest Neel temperature $\rm T_N=155$ K. Experimental measurements \cite{wildes2015magnetic, wildes2022magnetic,kim2019suppression,kim2021magnetic,kim2018charge,scheie2023spin} and DFT calculations \cite{chittari2016electronic, lane2020thickness,ushakov2013magnetism,wildes2015magnetic,wildes2022magnetic,hwangbo2021highly,lanccon2018magnetic} show that below  $\rm T_N$ Ni spins form a zig-zag antiferromagnetic ground state featured as double parallel ferromagnetic chains antiferromagnetically coupled within the single layer (see Fig.\ref{fig:f1}). Large spacing between adjacent layers suppresses interlayer exchange such that the antiferromagnetic order acquires a 2D character even in the bulk form \cite{wildes2015magnetic}. Spin dynamics in $\rm NiPS_3$ has been proved by high-resolution spectroscopy methods \cite{brec1986review, wildes2015magnetic}, and linear spin-wave theory using a Heisenberg Hamiltonian with single-ion anisotropies was applied to determine the magnetic exchange parameters and the nature of the anisotropy in $\rm NiPS_3$ samples \cite{olsen2021magnetic,kim2021magnetic,lanccon2018magnetic}.
In-plane magnetic exchange interactions up to third-nearest neighbors were required to account for the results. The nearest-neighbor exchange was found ferromagnetic with $\rm J_1\sim 2.5$ meV and the dominant antiferromagnetic third-neighbor exchange $\rm J_3\sim 13$ meV. Both, $\rm XY-$ like anisotropy and a small uniaxial component were required to fit the experimental results which leaded to two low-energy spin wave modes appearing in the spin-wave spectrum at the Brillouin zone center \cite{wildes2022magnetic}. The anisotropic Heisenberg Hamiltonian with up to three nearest neighbor couplings could reproduce the spin-wave energies but was at odds with the calculated neutron intensities showing that the classical spin models accounting for its magnetism up to date are a subject of debate. \cite{chandrasekharan1994magnetism, kim2019suppression,lane2020thickness,wildes2015magnetic,kim2021magnetic, chittari2016electronic}. Further, the presence of orbital degeneracy combined with the small magnitude of Ni spins suggests that quantum aspects could play a role in the magnetic properties of $\rm NiPS_3$. 

Aimed to find the spin model responsible for the magnetism in $\rm NiPS_3$, here we study the electron exchange mechanisms of a multi-band Hubbard model for the Ni atoms in $\rm NiPS_3$ in the limit of strong Coulomb interactions. d orbitals in transition-metals are localized, and direct exchange hopping can only occur between orbitals on different atoms that are very close to each other \cite{harrison2012electronic}, which makes direct hopping unlikely in $\rm NiPS_3$. Therefore the exchange mechanisms are extended by taking into account hopping via intermediate p orbitals located at the sulfur atoms in between two Ni sites. Integrating out the high energy states of the microscopic model,  spin exchanges were computed. Besides ferromagnetic and antiferromagnetic bilinear spin interactions, we found that a ferromagnetic biquadratic spin coupling is important in  $\rm NiPS_3$, giving rise to the following bilinear-biquadratic effective spin Hamiltonian for the Ni atoms:
\begin{eqnarray}
\mathcal{H}&=&-J_1\sum_{<ij>}({\bf S}_i\cdot{\bf S}_j)+J_3\sum_{(ik)}({\bf S}_i\cdot{\bf S}_k)\nonumber\\&& -\,K_1\sum_{<ij>}({\bf S}_i\cdot{\bf S}_j)^2
\label{eq:H}
\end{eqnarray}
where $\rm J_1$ and $\rm J_3$ denote the first and third nearest neighbor spin exchange couplings, respectively, and $\rm K_1$ is the first neighbor biquadratic spin exchange. The computed spin couplings are shown in Table \ref{table:1} and were used to evaluate the system's variational ground state energy considering the quantum nature of spins in $\rm NiPS_3$. We found that the zig-zag magnetic order corresponds to the ground state of Eq.\ref{eq:H} in the relevant space of parameters for bulk $\rm NiPS_3$ \cite{lanccon2018magnetic}. The zig-zag pattern coexists with a ferroquadrupolar order and competes with magnetic commensurate and incommensurate magnetic spirals. In monolayer $\rm NiPS_3$,  trigonal distortions \cite{kim2019suppression} found in experimental samples, change $\rm J_1$ to smaller and antiferromagnetic values which could drive monolayers to a Neel magnetic phase \cite{lanccon2018magnetic}.  
\begin{figure}
\includegraphics[width=\columnwidth]{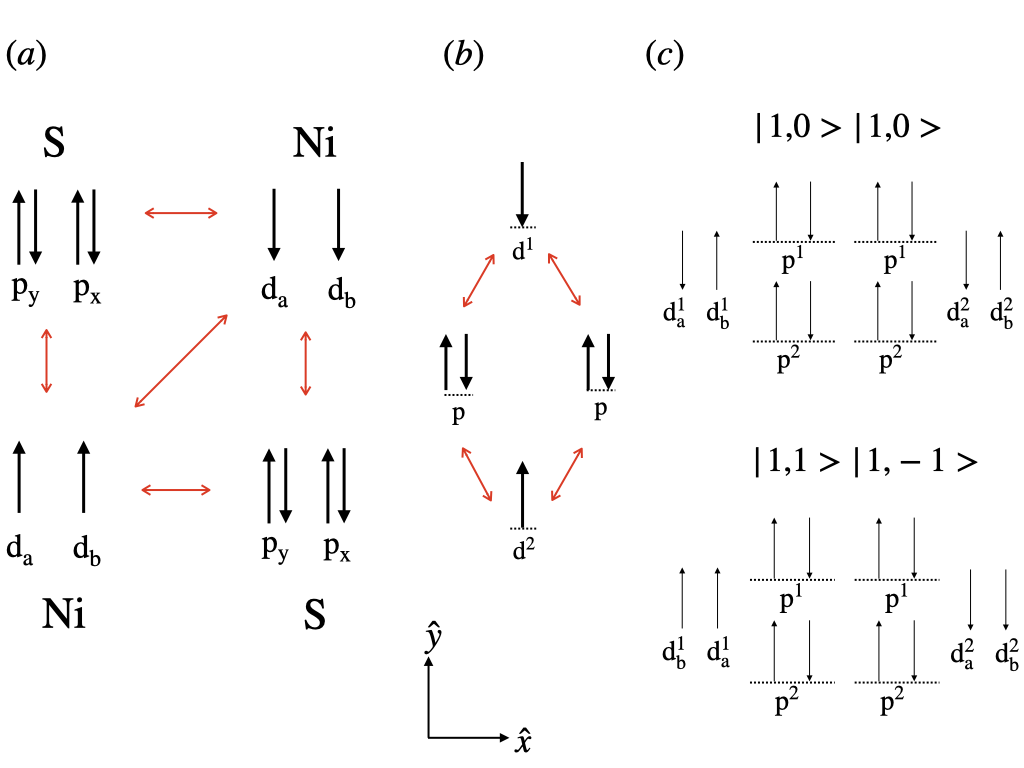}
    \caption{Proposed superexchange mechanism between (a) nearest neighbors d  orbitals in $\rm NiPS_3$ and (b) third nearest neighbors. (c) Two initial spin states with total spin S=1.}
    \label{fig:f2}
\end{figure}

In bulk $\rm NiPS_3$ the octahedral crystal field at Ni sites causes the 3d orbitals to split into a triplet of lower energy, $d_{xy}$, $d_{xz}, d_{yz}$ and a doublet $d_{x^2-y^2}, d_{3z^2-r^2}$, Fig\ref{fig:f1}(b) \cite{brec1986review,sugano2012multiplets,chang2022trigonal,koch2012exchange}. Each $\rm Ni^{2+}$ is in a $\rm d^8$ electronic configuration. Consequently, d orbitals belonging to the triplet are fully occupied while the doublet is half filled. While bulk samples have a monoclinic structure with point group $C_{2h}$, monolayers have a hexagonal structure with point group $D_{3d}$ \cite{kim2018charge}. In monolayers, due to trigonal distortions \cite{kim2019suppression} the crystal field splits d orbitals into the fully occupied triplet $d_{x^2-y^2}$, $d_{3z^2-y^2}, d_{xy}$, and the doublet $d_{xz}, d_{yz}$ \cite{kertesz1984octahedral,autieri2022limited,wilson1969transition,jiang2021recent}.
 Consider the case of two atoms of Ni with two d orbitals, each forming a spin S=1 state, and two atoms of sulfur symmetrically located in between the Ni and with two fully occupied p orbitals, Fig.\ref{fig:f2}(a). Operator $c_{i\alpha\sigma}^{\dagger}$ creates a spin electron $\sigma=\uparrow,\downarrow$  in the $\alpha$ orbital at site $i$ and $n_{i,\alpha,\sigma}=c_{i\alpha\sigma}^{\dagger}c_{i\alpha\sigma}$ defines the number of electrons at site i and orbital $\alpha$ with spin $\sigma$. The difference in energy of p and d orbitals is denoted $\Delta$ \cite{sugano2012multiplets,mattis2012theory}. Onsite Coulomb repulsion of two electrons in a single d orbital is $U_d$; electron repulsion in p orbitals is neglected. The transfer of one electron from p to d orbital has associated the energy  $U=U_d+\Delta$. $t_{\alpha\beta}^{ij}$ denotes the hopping amplitude from orbital $\alpha$ at site i to orbital $\beta$ at site j. Intrasite orbital hopping cancels since orbitals at the same site are orthogonal. States of interest have a fixed magnetic moment $S=1$ per site (atom). Interband charge interactions are considered, therefore Hund's couplings $J_H$ and $J_{H}^{'}$ are included at Ni and sulfur sites respectively \cite{koch2012exchange}.    
Altogether the microscopic model \cite{klein1973perturbation,van1994extended} becomes:
 \begin{eqnarray}
  H&=&\sum_{\substack{i\neq j\\ \alpha\neq\beta\\\sigma}}t_{\alpha\beta}^{ij}(c_{i\alpha\sigma}^\dagger c_{j\beta\sigma}+H.c.)\nonumber\\&& +\,U\sum_{i,\alpha} \nonumber n_{i\alpha\uparrow}n_{i\alpha\downarrow}-J_H\sum_{i,\sigma,\alpha\neq \beta} n_{i\alpha\sigma}n_{i\beta\sigma} 
  \\&& -\, J_H^{'}\sum_{i,\alpha\neq \beta} (c_{i\alpha\uparrow}^{\dagger}c_{i\alpha\downarrow}c_{i\beta\downarrow}^{\dagger}c_{i\beta\uparrow}+H.c.)  
\label{eq:Hm}
\end{eqnarray}
 In $\rm NiPS_3$ the spin exchange at the nearest neighbor level results from a competition between direct overlap of d orbitals and indirect hopping mediated by sulfur atoms \cite{koo2021unusual}. In the direct process, electrons hop between Ni orbitals at different sites of the honeycomb lattice. The indirect exchange mechanism known as superexchange \cite{anderson1950antiferromagnetism,koch2012exchange,mila2000origin} is mediated by the virtual hopping to two sulfur ions in between the two Ni atoms. This is a more realistic situation for $\rm NiPS_3$ because in transition metals compounds, the overlap between d orbitals \cite{harrison2012electronic} separated a distance r decays as $\sim 1/r^5$. Consider the two initial states of Fig.\ref{fig:f2}(c) where spins at different Ni sites are antiparallel. They are denoted $\ket{1,0}\ket{1,0}$ and $\ket{1,1}\ket{1,-1}$, according to the notation $\ket{S^1,m^1}\ket{S^2,m^2}$ where $S^k$ and $m^k$ represent respectively the total spin quantum number of site k and the z component of the total spin.
With four available half-filled d-orbitals, up to four electrons could hop. The electron hopping from the initial states with $\rm S^k=1$ gives rise to intermediate states with one, two, three, and four double-occupied d-orbitals where $\rm S^k\neq 1$. Indirect interactions across intermediate states  mediate interactions between $\ket{1,0}\ket{1,0}$ and $\ket{1,1}\ket{1,-1}$. To integrate out high energy states, the electron Hamiltonian matrix from Eq.\ref{eq:Hm} was separated into $\rm H=\tilde{H}_{0,0}+\tilde{H}+\tilde{T}$. $\rm\tilde{H}_{0,0}$ and $\rm\tilde{H}$ contain the on-site contributions due to Coulomb interactions and interband charge interactions of the low-energy states with no double occupied d orbitals, ($\rm\tilde{H}_{0,0}$), and the high energy states with at least one double occupied d-orbital ($\rm\tilde{H}$). $\rm\tilde{T}$ contains off-diagonal terms due to electron hopping (details in  Supplementary Material). $\rm H$ was partitioned into blocks where diagonal matrices contain the energy of the basis states with k double occupied d-orbitals ($\rm\tilde{H}_{k,k}$). Off diagonal blocks are hopping matrices $\rm\tilde{T}_{k-1,k}$ that connect states with $k-1$ and $k$ double occupied d-orbitals:
\[
H=\begin{pmatrix}
\tilde{H}_{0,0} & \tilde{T}_{0,1} & \tilde{0}& \tilde{0} & \tilde{0}\\
\tilde{T}_{1,0} & \tilde{H}_{1,1} & \tilde{T}_{1,2}& \tilde{0} & \tilde{0}\\
\tilde{0} & \tilde{T}_{2,1} & ...& ... & ...\\
\tilde{0} & \tilde{0} & ...& \tilde{H}_{k-1,k-1} & \tilde{T}_{k-1,k}\\
\tilde{0} & \tilde{0} &...& \tilde{T}_{k,k-1} & \tilde{H}_{k,k}
\end{pmatrix}
\]
\begin{table}[bt]
\begin{tabular}{|c||c|}
\hline
$J_1^s$ & $- 
\rm t_{pd}^4\frac{J_{H}'(J_H-2 U)^2}{2 U^4 (J_H-U)^2}$\\
\hline
$\rm J_1^D$ & $\rm t_{dd}^2\frac{2(J_H-2 U_d)}{U (J_H-U_d)}$\\
\hline
$\rm J_3$ & $\rm t_{pd}^4\left(\frac{1}{U_d}+\frac{1}{(U_d+\Delta)}\right)\rm\frac{4}{(U_d+\Delta)^2}$\\
\hline
$\rm K_1$ &$\rm-2J_{H}'\left[t_{pd}^6\frac{ \left(16 J_H^2-63 J_H U+72 U^2\right)}{18 U^6
   (J_H-U)^2}+t_{pd}^8\frac{(47 J_H-108 U) (5 J_H-12 U)}{864 U^8 (J_H-U)^2}\right]$\\
\hline
\end{tabular}
\caption{Spin couplings for spins in Ni atoms of $\rm NiPS_3$ computed from direct and superexchange processes in Eq.\ref{eq:Hm}.}
\label{table:1}
\end{table}
\begin{table}[bt]
\begin{tabular}{ |c||c|c| }
\hline
$\rm NiPS_3$&Monolayer&Bulk\\
\hline
\hline
{$\rm t_{dd}$}
	& $\rm t_{d_{xz}d_{xz}}\sim t_{dd\pi}$ & $\rm t_{d_{x^2-y^2}d_{x^2-y^2}}\sim t_{dd\pi}$\\ 
& $\rm t_{d_{yz}d_{yz}}\sim t_{dd\pi}$& $\rm t_{d_{3z^2-z^2}d_{3z^2-r^2}}\sim \frac{1}{4}t_{dd\sigma}$\\ 
\hline
{$\rm t_{pd}$}
		& $\rm t_{d_{xz}p_{x}}\sim t_{pd\pi}$ & $\rm t_{d_{x^2-y^2}p_{x}}\sim \frac{\sqrt{3}}{2}t_{pd\sigma} $\\ 
& $\rm t_{d_{yz}p_{x}}\sim t_{pd\pi}$ & $\rm t_{d_{3z^2-r^2}p_{x}}\sim \frac{1}{2}t_{pd\sigma}$\\ 
&$\rm t_{d_{xz}p_{y}}\sim t_{pd\pi}$& $\rm t_{d_{x^2-y^2}p_{y}}\sim \frac{\sqrt{3}}{2}t_{pd\sigma}$ \\ 
&$\rm t_{d_{yz}p_{y}}\sim t_{pd\pi}$ & $\rm t_{d_{3z^2-r^2}p_{y}}\sim \frac{1}{2}t_{pd\sigma}$\\ 	
\hline
$\rm J_1^D$ 
	& $5\times 10^{-4}$
	& $5\times 10^{-4}$ 
	\\
\hline
$\rm J_1^s$ 
	& $-2\times 10^{-4}$
	&$-2\times 10^{-3}$ 
 	\\
 \hline
$\rm J_1=J_1^D+J_1^s $
	& $3\times 10^{-4}$
	& $-1\times 10^{-3}$
	\\
\hline
$\rm K_1/J_1 $
	& $-7\times 10^{-3}$
	&$ 1\times 10^{-2}$ \\
\hline
$\rm J_3/J_1 $
	& 1.2 
	& -3
	\\
 \hline
\end{tabular}
\caption{Slater-Koster integrals between d and between p and d-orbitals from ref.\cite{harrison2012electronic}. Approximate numerical evaluation (in [eV] units) of the spin couplings presented in Table \ref{table:1} in bulk and monolayer of $\rm NiPS_3$  using parameters for $\rm NiPS_3$ from the literature \cite{lane2020thickness,lanccon2018magnetic,autieri2022limited} such as U=6 [eV] $\rm J_H=J_H^{'}=0.7$ [eV], $\rm \Delta=3$ [eV], $\rm t_{dd}=0.05$ [eV], $\rm t_{pd\sigma}=1$ [eV].}
\label{table:2}
\end{table}

The subspaces are decoupled through a canonical transformation using the perturbative approach of L{\"o}wdin \cite{lowdin1951note} and Schrieffer-Wolff \cite{schrieffer1966relation,bravyi2011schrieffer} where $\rm\tilde{H}$ and $\rm\tilde{T}$ are treated as perturbation (see Supplementary Material). In this way, high energy states are down-folded into the energetically well-separated sector of interacting spins of constant quantum number at each site \cite{koch2012exchange}. This is a consequence of Hund's coupling and onsite Coulomb interactions, which are large with respect to the hopping amplitudes in transition metal compounds \cite{hoffmann2020systematic}.
For nearest neighbor Ni atoms (1-nn), sulfur ions form a ninety degrees bridge between the two Ni sites,  Fig.\ref{fig:f1}(a) and Fig.\ref{fig:f2}(a) \cite{harrison2012electronic,kertesz1984octahedral,koch2012exchange}. By symmetry, there is only hopping between d and p orbitals that point to each other. Therefore in the superexchange process at 1-nn level at each Ni site, one of the d-orbitals could overlap with the $p_x$ orbital of one of the sulfur atoms, and the other could overlap with the $p_y$ orbital of the second, Fig.\ref{fig:f2}(a). The superexchange Hamiltonian matrix contains five diagonal blocks with the energy of:  the two initial states, $\rm\tilde{H}_{0,0}$, the eight excited states with one double occupied d-orbital, $\rm\tilde{H}_{1,1}$, the twelve states with two double occupied d-orbitals, $\rm\tilde{H}_{2,2}$, the eight with three double occupied d-orbitals, $\rm\tilde{H}_{3,3}$, and the two excited states with four double occupied d-orbitals, $\rm\tilde{H}_{4,4}$. The off diagonal matrix elements consist of hopping matrices ($\rm\tilde{T}_{0,1}$, $\rm\tilde{T}_{1,2}$, $\rm\tilde{T}_{2,3}$, $\rm\tilde{T}_{3,4}$) between the basis states. Thought expected to be small, direct exchange between 1-nn is also considered  here.
\begin{figure}
\includegraphics[width=\columnwidth]{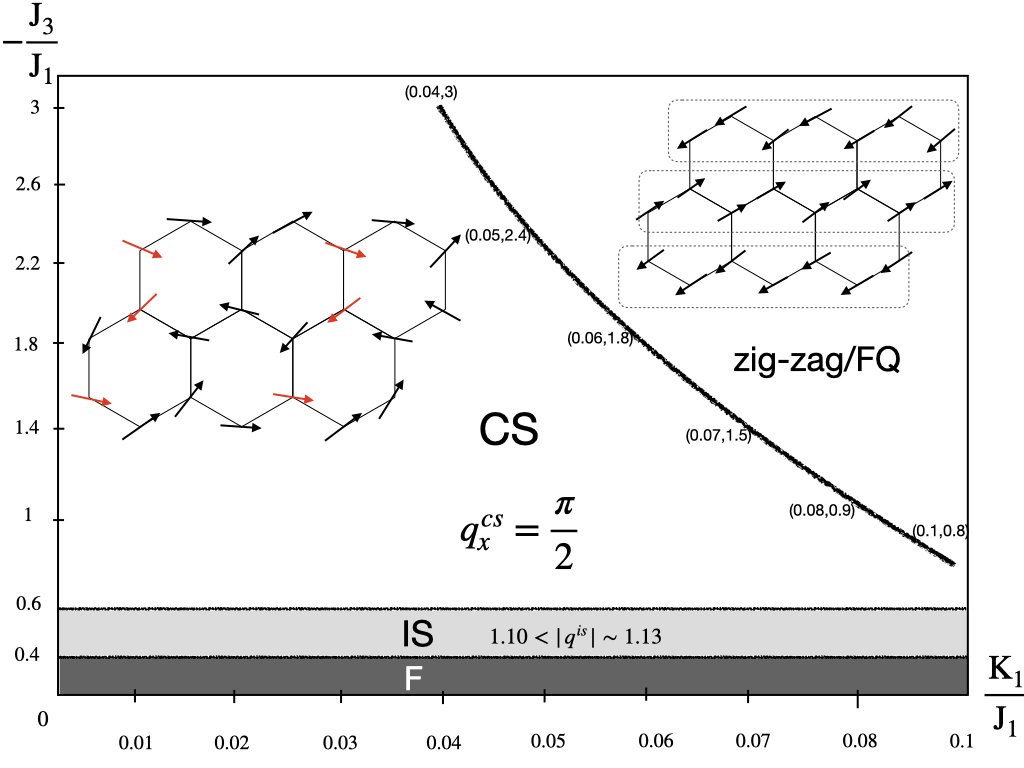}
    \caption{Variational phase diagram of Eq.\ref{eq:H}, at T=0. Arrows in red illustrate periodicity of spins in the spiral phase CS.}
    \label{fig:f3}
\end{figure}
 Second neighbors Ni atoms (2-nn) do not share a common sulfur ion in $\rm NiPS_3$, (Fig.\ref{fig:f1}); therefore, the electron exchange proceeds via direct overlap, which for 2-nn is neglected. 
 In the case of third nearest neighbor Ni atoms (3-nn), the angle between two 3-nn Ni and the two sulfur ions in between is larger than ninety degrees, Fig.\ref{fig:f1}. Consequently, superexchange between two 3-nn Ni could be mediated by a single p orbital \cite{koch2012exchange}, which serves the two Ni ions, Fig.\ref{fig:f2}(b). At 3-nn level, direct exchange is neglected and we only consider bilinear spin exchanges; therefore, only virtual hopping of one and two electrons are computed. 

Downfolding the high energy states by going up to fourth order in perturbation theory \cite{lowdin1951note,schrieffer1966relation,bravyi2011schrieffer} and by writing spin operators in terms of electron operators  $\rm c_{i\alpha\sigma},c_{i\alpha\sigma}^\dagger$ yields the effective spin Hamiltonian of the electron system (as shown in the Supplementary Material). Mediated by two orthogonal p orbitals, the spin couplings $\rm J_1^s<0$ and $\rm K_1<0$ are ferromagnetic, and they correspond respectively to bilinear and biquadratic 1-nn spin interactions originated from four, six and eight virtual hoppings. $\rm J_1^D$ originates from a direct exchange between Ni atoms and is antiferromagnetic. $\rm J_3>0$ originates from four virtual hopping mediated by a single p orbital and consequently is antiferromagnetic \cite{koch2012exchange,mila2000origin}. Explicit expressions for $\rm J_1=J_1^s+J_1^D$, $\rm J_3$ and $\rm K_1$ are shown in Table \ref{table:1}. There,   
terms $\rm{\mathcal{O}}(t_{pd}^4)$ arise from two double occupied d-orbitals. Terms proportional to $\rm{\mathcal{O}}(t_{pd}^6)$ correspond to three double occupied d orbitals from six p-d hopping processes, and the term $ {\mathcal{O}}(t_{pd}^8)$ accounts for four double occupied d-orbitals. Terms proportional to ${\mathcal{O}}(t_{dd}^2)$ are due to one doubly occupied d-orbital due to direct exchange.

While in bulk $\rm NiPS_3$ the octahedral crystal field favors the doublet $\rm d_{x^2-y^2}, d_{3z^2-r^2}$,  in monolayers the trigonal distortion of such a field \cite{kim2019suppression} favors the doublet $\rm d_{xz}, d_{yz}$, Fig.\ref{fig:f1}(b) \cite{wilson1969transition,koch2012exchange,harrison2012electronic}. Even though the trigonal distortion could change the angles between Ni and S atoms \cite{harrison2012electronic,koch2012exchange,wilson1969transition} here we neglect this important effect and assume that the different symmetry of the participating d-orbitals in bulk and monolayer does not change the superexchange mechanism presented above; however, it affects the values of the hopping integrals according to the Slater-Koster scheme \cite{ slater1954simplified,harrison2012electronic} as shown in Table \ref{table:2}. 
 Taking into account the Slater-Koster rules, considering all possible combinations of d and p orbital hopping \cite{harrison2012electronic} and using values of U, $\Delta$ and $\rm J_H$ from the literature \cite{autieri2022limited} the spin couplings from Table \ref{table:1} in bulk and monolayer $\rm NiPS_3$ are evaluated and shown in Table \ref{table:2}. We find that in the case of bulk samples $\rm |J_1^D|<|J_1^s|$.
 
To study magnetic orders of Eq.\ref{eq:H} at T=0, we consider the trial ground state $ \ket{\psi}=\otimes_i\ket{\psi_i}$ \cite{lauchli2006quadrupolar,ivanov2003effective,mattis2012theory}. It consists of an entanglement-free direct product of arbitrary wavefunctions with spin S=1 at each site \cite{lauchli2006quadrupolar,stoudenmire2009quadrupolar}. A general single spin state can be written as the coherent state 
\begin{equation}
\ket{\psi_i}=b_{ix}\ket{x}+b_{iy}\ket{y}+b_{iz}\ket{z}    
\end{equation}
where $\bf{b}_i$ is an arbitrary complex vector satisfying the normalization constraint $\bf{b}_i^*\cdot \bf{b}_i=1$, and we have chosen the time-reversal invariant basis of the $SU(3)$ (S=1) fundamental representation
$\ket{x}=\frac{i\ket{1}-i\ket{\tilde{1}}}{\sqrt{2}}\text{ , }\quad   \ket{y}=\frac{i\ket{1}+i\ket{\tilde{1}}}{\sqrt{2}}\text{ , } \ket{z}=-i\ket{0}$
where $\ket{1},\ket{\tilde{1}}$ and $\ket{0}$ are the three cartesian spin-1 states quantized along the z axis \cite{ivanov2003effective}. The basis states satisfy  $S^x\ket{x}=S^y\ket{y}=S^z\ket{z}=0$.
The magnetization of the system is defined through the expectation value of the spin at each site \cite{stoudenmire2009quadrupolar}
\begin{equation}
 M=\sum_k\bra{\psi_k}{\bf S}_k\ket{\psi_k}=-i\sum_k{\bf b}_k^*\times {\bf b}_k   \end{equation}
In terms of the complex vectors $\bf b_i$ the expectation value of the $\rm J_1-J_3-K_1$ spin Hamiltonian of Eq.\ref{eq:H}  becomes 
\begin{eqnarray}
\bra{\psi}\mathcal{H}\ket{\psi}&=&\sum_{<ij>}\left[-J_1|{\bf b_i^*}\cdot {\bf b_j}|^2+(J_1-K_1)|\bf b_i\cdot b_j|^2\right]\nonumber\\&& +\,J_3\sum_{(ij)}\left[|\bf b_i^*\cdot b_j|^2-|\bf b_i\cdot b_j|^2\right]
\label{eq:Hq}
\end{eqnarray}
Because of the biquadratic interaction, we investigate a possible quadrupolar order, QP in the system \cite{stoudenmire2009quadrupolar}. To that purpose we introduce the quadrupolar operator QP, a tensor with five components ${\bf Q_i}=[Q_i^{x^2-y^2},Q_i^{3z^2-r^2},Q_i^{xy},Q_i^{yz},Q_i^{xz}]=[(S_i^x)^2-(S_i^y)^2,(2(S_i^z)^2-(S_i^x)^2-(S_i^y)^2)/\sqrt{3},S_i^xS_i^y+S_i^yS_i^x,S_i^yS_i^z+S_i^zS_i^y,S_i^xS_i^z+S_i^zS_i^x]$. QP can be expressed in terms of the complex $\bf b$ vectors, 
\begin{equation}
\bra{\psi_i}{\bf Q}_{i\mu,\nu}\ket{\psi_i}=\frac{1}{3}\delta_{\mu,\nu}-\frac{1}{2}(b_{i\mu}^*b_{i\nu}+b_{i\nu}^*b_{i\mu})    
\end{equation} Using the identity $\rm{\bf Q_i}\cdot {\bf Q_j}=2({\bf S_i}\cdot {\bf S_j})^2+{\bf S_i}\cdot {\bf S_j}-2/3$, in terms of the QP operators the $\rm J_1-J_3-K_1$ Hamiltonian, Eq.\ref{eq:H} can be written as:
\begin{eqnarray}
\mathcal{H}&=&-(J_1-\frac{K_1}{2})\sum_{<ij>}({\bf S}_i\cdot{\bf S}_j)
+J_3\sum_{(ik)}({\bf S}_i\cdot{\bf S}_k)\nonumber\\&& -\,\frac{K_1}{2}\sum_{<ij>}({\bf Q}_i\cdot{\bf Q}_j)-\sum_{i}\frac{4}{3}K  
\label{eq:HQ}
\end{eqnarray}
To find variational ground states of the spin Hamiltonian Eq.\ref{eq:H}, the variational Eq.\ref{eq:Hq} was minimized respect to all complex  ${\bf b_i}$ vectors in a hexagonal cluster of twenty-two spins (as the one shown in the inset of Fig.\ref{fig:f3}) by using the \textit{Nminimize} procedure from Wolfram Mathematica \cite{wolf} (see Supplementary Material). Fig.\ref{fig:f3} presents the corresponding $\rm -J_3/J_1$ v/s $\rm K_1/J_1$  phase diagram of Eq.\ref{eq:H} at T=0. We considered ferromagnetic nearest neighbour couplings and antiferromagnetic $\rm J_3$ in the range of results valid for bulk samples, given in Tables \ref{table:1} and \ref{table:2}. 
Four magnetic phases and one quadrupolar order have been identified. 
For all values of $\rm K_1$ and for $\rm J_3$ in the range $\rm0\leq|J_3|\leq0.4J_1$ the system settles in a ferromagnetic state F, but as $\rm J_3$ enters in the range $\rm0.4J_1\leq|J_3|\leq0.6J_1$ spins rearrange into incommensurate spirals IS out of the $x-y$ plane. For $\rm J_3\geq 0.6J_1$ two magnetic phases can arise. At $\rm K_1>0.4J_1$ a zig-zag phase with spins canted out of the $x-y$ plane competes with a commensurate spiral magnetic order CS with wavevector $\rm q_{x}^{cs}=\frac{2\pi}{3}$. The zig-zag phase with zero average spin moment coexists with a uniaxial ferroquadrupolar order FQ \cite{stoudenmire2009quadrupolar}. The CS order (inset of Fig.\ref{fig:f3}) is a noncoplanar spiral where generally spins settle out of the $x-y$ plane and give rise to non zero QP moments \cite{stoudenmire2009quadrupolar,lauchli2006quadrupolar}. 
Magnetic phases and quadrupolar order were identified by inspecting the ground state spin textures obtained from the variational results and were confirmed by computing the space correlation functions through the static structure factor of the magnetization and quadrupolar order parameters C, $\rm S({\bf q})=e^{i{\bf r}_{ij}\cdot{\bf q}}\bf\langle C_i\cdot C_j\rangle$ for $q\in (0,\pi)$ (as shown in the Supplementary Material). Phase boundaries were identified by computing the second derivative of the ground state energy, with respect to the couplings $\rm J_3$ and $\rm K_1$, and looking for singular features indicative of changes in the ground state phases. 

Quadrupolar and spin correlations grow toward larger values of $\rm K_1/J_1$, and from IS, toward larger values of $\rm |J_3/J_1|$.
Inspecting  Eq.\ref{eq:HQ}, in the case of $\rm J_3=0$ a ferromagnetic ground state is expected as long as $|J_1|>\frac{|K_1|}{2}$. Variational results show that increasing $\rm K_1$ drives the spins in CS toward a coplanar spiral phase until the system reaches the collinear zig-zag state. 

For $\rm J_3\geq 0$ and $\rm(J_1-\frac{K_1}{2})>0$ magnetic frustration could lead the system to disordered phases. But if $\rm J_3$ is large enough, the zig-zag state is favoured where a single spin is arranged ferromagnetically with two of the nearest neighbors in its chain and antiferromagnetically with the third one in a next neighbor chain. Nevertheless, the zig-zag state becomes more difficult to accomplish as $\rm J_3$ and $\rm K_1$ approach the limit $\rm |J_1-\frac{K_1}{2}|\sim |J_3|$. In this case our variational calculations show that the IS phase arises.

To further investigate the spiral order in bulk $\rm NiPS_3$, we introduced the anzatzs ${\bf b_i}=(\rm {\sin\theta\cos({{\bf q}\cdot{\bf r}_i})e^{i\phi},\sin\theta\sin({{\bf q}\cdot{\bf r}_i})e^{i\phi},\cos{\theta}})$ \cite{stoudenmire2009quadrupolar} for the vector of amplitudes $\bf b_i$ describing the wavefunction at a single site located at position $\bf r_i$ in the honeycomb lattice. 
Now Eq.\ref{eq:Hq} was minimized with respect to ${\bf q},\theta,\phi$ in the hexagonal cluster with twenty-two spins introduced before, and the results were compared with the variational ones confirming zig-zag, F, and QP orders, as well as the wavevector $\rm{\bf{q_x}}^{cs}=2\pi/3$. In the range of parameters of IS we found the order wavevector $1.10<|{\bf{q}}^{is}|\sim 1.15$.

Previous studies have reported that the orientation of magnetic moments in $\rm NiPS_3$ could be influenced by biaxial magnetocrystalline anisotropy consisting of a dominant easy-plane anisotropy that locks the orientation of the spins to a magnetic plane $x-y$ slightly inclined from the crystallographic ab plane, and a secondary weaker anisotropy that orients the spins in the magnetic $x-y$ plane along the x-axis \cite{lanccon2018magnetic,kim2021magnetic}. To determine its effects on the phase diagram of Fig.\ref{fig:f3} the term $ A\sum_{j}(S_j^z)^2=A\sum_{j}\left[i\bf (b_j^*\times b_j)\cdot\hat{z}\right]^2$ weas added to Eq.\ref{eq:H}, where A plays the role of the anisotropic coupling. Variational ground states were subsequently computed for the range of parameters of Fig.\ref{fig:f3} and $0\leq A\leq 2\times 10^{-1}J_1$ according to reported values of A \cite{autieri2022limited,olsen2021magnetic,lane2020thickness}. The resulting variational magnetic phases coincide with the ones of Fig.\ref{fig:f3}. Now F, zig-zag,  CS, and IS settle near the $x-y$ plane, and the QP phase develops a uniaxial director vector along the $\hat{z}$ axis. At $A\sim 1.4\times 10^{-1}J_1$, the zig-zag phase settles in the $x-y$ plane, slightly inclined out of it by an angle $\sim 8$ degrees. The boundary between zig-zag and spiral phases moved slightly toward smaller $\rm K_1$ and $\rm J_3$ after the inclusion of easy plane anisotropy (as shown in the Supplementary Material).

 Results from Monte Carlo simulations of an anisotropic Heisenberg model with first and third nearest neighbor interactions using ab-initio parameters support the hypothesis that a Neel order could compete with the zig-zag in $\rm NiPS_3$ \cite{lane2020thickness}. That could be the case if the nearest neighbour couplings $J_1$ were positive or if $J_1<\frac{|K_1|}{2}$,  which based in our calculations is not possible for this material: Table \ref{table:2} shows $\rm K_1$ to be about one order of magnitude smaller than $\rm J_1$ in bulk; thus, a Neel state is unlikely to occur, at least in bulk $\rm NiPS_3$. 
 However for the case of monolayers where $\rm J_1$ is positive, and $K_1$ is negative and much smaller than $\rm J_1$, the variational calculations confirm a Neel order (Supplementary Material). 
 
Depending on the relative strength between the spin couplings, Eq.\ref{eq:H} gives rise to four magnetic phases at T=0. The zig-zag phase, which has been found in bulk samples of $\rm NiPS_3$, is the most likely to occur and competes with spiral magnetic patterns when we use the available values for Coulomb repulsion, Hund's coupling and d-p gaps from DFT calculations in the computed bulk spin exchange couplings. One aspect that has remained controversial is whether or not the zig-zag order survives up to the monolayer limit \cite{lane2020thickness,kim2021magnetic}. Experiments suggest that the crystal field of bulk and monolayer $\rm NiPS_3$ differs \cite{kim2019suppression,hempel1981ground,chang2022trigonal} and here we assume that a consequence is that the active d orbitals in both cases have a different symmetry \cite{wilson1969transition,koch2012exchange, autieri2022limited,harrison2012electronic}. Assuming that the angle between Ni and S atoms does not change due to the trigonal distortion in monolayers, we have shown that a possible consequence of that is the relative decrease of the spin couplings in monolayer systems. The positive and very small $\rm J_1$ of monolayers (Table \ref{table:2}) could drive samples toward a Neel phase, an IS  or a disordered state. Which path monolayers will take it is unknown at this point: for such small $\rm J_1$,  quantum fluctuations or interactions like second nearest neighbors could play a role and effects like disorder could become relevant \cite{stoudenmire2009quadrupolar}. One important aspect not considered here is the deviation of the ninety degrees angle between Ni and sulfur atoms in monolayers \cite{brec1986review}. According to Goodenough-Kanamori rules  \cite{goodenough1955theory,kanamori1959superexchange}, such deviations could favor a superexchange mediated by a single p orbital, which would increase the magnitude of the antiferromagnetic $\rm J_1$  in monolayer samples. If that were the case, the variational calculations show that a Neel order is favored in monolayers. 
\section*{Supplementary Material}
It contains details of calculations of the microscopic Hamiltonian and the perturbative approach to find the effective spin model. It also includes additional figures of variational spin correlations and magnetic phases.
\begin{acknowledgments}
This work was supported by Fondecyt under Grant No. 1210083. The author thanks Je-Geun Park, Joerg Schmalian, and  Alexander Mirlin for valuable discussions.
\end{acknowledgments}

\section*{Data Availability Statement}
The data that support the findings of this study are available within the article [and its supplementary material].

\bibliography{v0}% Produces the bibliography via BibTeX.

\end{document}